# Superconducting nanowire single photon detector at 532 nm and demonstration in satellite laser ranging

Hao Li,[1] Sijing Chen,[1] Lixing You,[1,3] Wengdong Meng,[2] Zhibo Wu,[2] Zhongping Zhang,[2,4] Kai Tang,[2] Lu Zhang,[1] Weijun Zhang,[1] Xiaoyan Yang,[1] Xiaoyu Liu,[1] Zhen Wang,[1] and Xiaoming Xie[1]

[1]*State Key Laboratory of Functional Materials for Informatics, Shanghai Institute of Microsystem and Information Technology (SIMIT), Chinese Academy of Sciences, 865 Changning Rd., Shanghai 200050, China*
[2]*Shanghai Astronomical Observatory, Chinese Academy of Sciences, Shanghai 200030, China;*
[3]*lxyou@mail.sim.ac.cn*
[4]*zzp@shao.ac.cn*

**Abstract**: Superconducting nanowire single-photon detectors (SNSPDs) at a wavelength of 532 nm were designed and fabricated aiming to satellite laser ranging (SLR) applications. The NbN SNSPDs were fabricated on one-dimensional photonic crystals with a sensitive-area diameter of 42 μm. The devices were coupled with multimode fiber ($\phi$ = 50 μm) and exhibited a maximum system detection efficiency of 75% at an extremely low dark count rate of <0.1 Hz. An SLR experiment using an SNSPD at a wavelength of 532 nm was successfully demonstrated. The results showed a depth ranging with a precision of ~8.0 mm for the target satellite LARES, which is ~3,000 km away from the ground ranging station at the Sheshan Observatory.

______________________________________________________________________________________________

## 1. Introduction

In the last decade superconducting nanowire single-photon detectors (SNSPDs) at near infrared wavelength of 1,550 nm have undergone significant improvement and received considerable attention due to their high detection efficiency (DE) [1-5], low timing jitter [6], high count rate[2, 5], and low dark count rate (DCR) [7-9], enabling numerous impressive applications such as long-distance quantum key distribution [10, 11], space-ground laser communication [12], depth imaging [13, 14], and on-chip characterization of nanophotonic circuits [5]. Indeed, SNSPDs performs excellently at shorter wavelength since photons of such wavelengths have larger photon energies, thereby increasing their detection probability. Besides, there are many other applications that need advanced single photon detectors (SPDs) at shorter wavelengths. There have been some reports on SNSPDs at shorter wavelengths that outperform semiconductor SPDs [15-18] and are useful for some specific applications such as fluorescence spectroscopy detection at 635 nm [18], or singlet oxygen luminescence detection at 1270 nm [19]. Satellite laser ranging (SLR), which is the most accurate technique currently available to determine the geocentric position of an Earth-orbiting satellite, contributes significantly to scientific studies on Earth/atmosphere/oceans system. The precision of SLR is highly dependent on the performance of SPDs, and most SLRs are operated at 532 nm using photomultipliers or avalanche photodiodes (APDs) [20, 21]. In the future, high performance SNSPDs may be good candidates for SLR applications at not only 532 nm but also at near infrared wavelengths (1,064 and 1,550 nm).

In this study, we fabricated SNSPDs at a wavelength of 532 nm for applications to SLR. To enhance the absorption at 532 nm, a specific substrate with a one-dimensional (1-D) photonic crystal (PC) comprising dielectric films is used. The sensitive area diameter was chosen to be 42 μm for ~99% coupling efficiency of the incident photons. The width and pitch of nanowire are chosen to be 140 nm and 280 nm, respectively. The device showed a maximal saturated system DE of 75% at DCR of <0.1 Hz. A preliminary demonstration of SLR using an SNSPD was performed. The experimental result showed a ranging resolution of 8 mm for the satellite LARES, which is ~3,000 km away from the ground-ranging station.

## 2. Device design

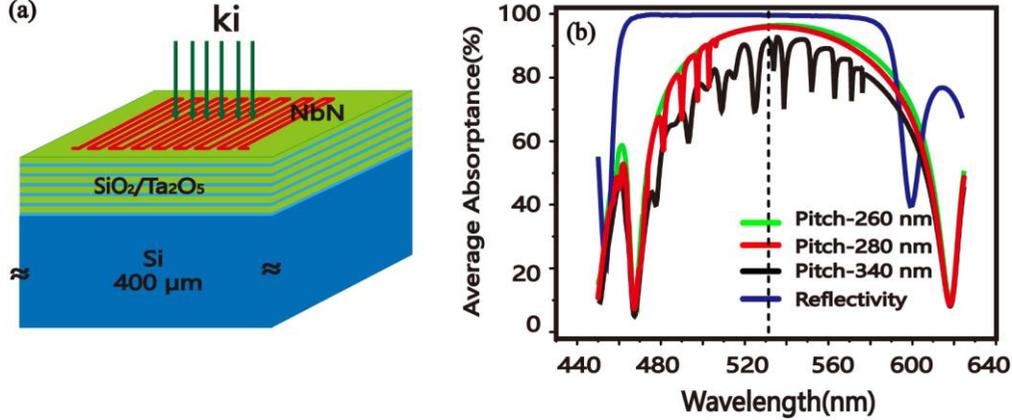

Fig. 1 (a) Schematic of the SNSPD based on a PC substrate. The PC structure was formed by multiple alternating $Ta_2O_5$ and $SiO_2$ layers on a Si substrate (blue). (b) The measured reflectivity (blue curves) of the PC substrate consisting of 13 bilayers of $Ta_2O_5$ and $SiO_2$. The other three curves illustrate the averaged absorptances of the nanowire for normal incidence with different pitches (260 nm, 280 nm and 340 nm) calculated using the RCWA method. The dips in the absorptance curves are caused by high order grating diffractions.

SNSPDs were designed based on a 1-D PC as shown in Fig. 1(a). The common optical film materials $Ta_2O_5$ and $SiO_2$ comprised the PC structure. The refractive indices were $n_{Ta2O5} = 2.15$ and $n_{SiO2} = 1.48$, respectively, which were measured using a spectroscopic ellipsometer at a wavelength of $\lambda_0 = 532$ nm at room temperature. The refractive index of NbN was $n_{NbN} = 3.37 + 3.64i$, at 532 nm, and the refractive index and thickness of the Si substrate were $n_{Si} = 4.14 + 0.04i$ and $l = 400$ μm, respectively. For a resonance absorptance around the target wavelength, the thicknesses of the layers were designed to be a quarter of the wavelength $\lambda_0$, $L_1 = 61.9$ nm for $Ta_2O_5$ and $L_2 = 89.9$ nm for $SiO_2$. Thirteen bilayers were used in our PC structure to obtain high reflectivity and acceptable fabrication complexity. Similarly, high absorption SNSPDs on the basis of PC structures were reported at other wavelengths [15-17, 22]. Unlike SNSPDs at longer wavelengths, both the width and pitch of the nanowire had to be optimized very carefully.

To quantitatively investigate the optical properties of the SNSPDs, we performed an electromagnetic (EM) simulation using the rigorous coupled-wave analysis (RCWA) method, in which the EM field in the periodic grating region (nanowire region) is expanded into a sum of spatial harmonics on the basis of Floquet's theorem. The polarization dependence of the SNSPDs may have limited the DE when the SNSPDs were coupled with a multimode fiber (MMF) because the polarization in MMFs is difficult to control due to the coupling of different propagating modes caused by fiber imperfections, such as index inhomogeneity, core ellipticity and eccentricity, and bends. Therefore, in the simulation, we define the average absorptance of our structure as $A = (A_{//} + A_{\perp})/2$, where $A_{//}$ and $A_{\perp}$ are the calculated absorptances for the parallel and perpendicular polarization plane waves, respectively.

To efficiently respond to visible photons at 532 nm, the width of the nanowire was designed to be 140 nm with a film thickness of 6.5 nm in accordance with previous experimental work on SNSPDs [15-17] and for the following reasons. Firstly, photons of 532 nm wavelength have a high photon energy, which can cause detection events in a wider nanowire with a higher probability. Secondly, a wider nanowire is insensitive to near-infrared photons, which make up the main portion of the room-temperature environmental blackbody radiation; as a result, wider nanowires may have fewer background dark counts. Thirdly, a wider nanowire may have high uniformity since the fabrication is easier.

Figure 1(b) presents the average absorptances of SNSPDs with different nanowire pitches. The dips in the absorptance curves are caused by diffraction losses which increase in quantity and shift toward the target wavelength with an increase in pitch value. For short pitches of SNSPDs, the nanowire works like a sub-wavelength grating where the diffraction losses are limited to only 0 order diffraction in the normal direction (for the normal incidence case)[23]. For larger pitches, higher-order diffractions in other directions arise and result in the absorption dips shown in Fig. 1(b). Thus, to avoid the dips around the target wavelength of 532 nm, the pitch should be small. On the other hand, we hope to increase the pitch of the nanowire as much as possible to reduce the kinetic inductance and ease the fabrication complexity. As a result, the pitch was finally determined to be 280 nm, giving a filling ratio of 50%.

For coupling of the photons to the SNSPD, the photons were directly guided to the SNSPD through a front-side aligned lensed MMF. The graded index lenses were spliced to the tip of the MMF with a 50-μm-diameter core. The incident light spot was focused to a minimal diameter, $2w$ of about 27 μm (the beam waist) where the device was located. To obtain a good coupling, the diameter $2r$ of the active area was selected to be $\pi w$ (~42 μm), corresponding to a coupling efficiency of 99% of a Gaussian beam.

## 3. Device fabrication and measurement

In our experiment, the 13 periodic $SiO_2/Ta_2O_5$ bilayers were alternately deposited onto a Si substrate using ion beam sputtering, with the film thickness optically monitored to ensure adherence to the designed layer thickness. The measured surface roughness of the PC was about 2 Å, measured by an atomic force microscope, and the measured reflectivity at 532 nm was more than 99 % as shown in Fig. 1(b) (blue curve). Then, an ultrathin (6.5 nm) NbN film was deposited on the PC substrate at room temperature using reactive DC magnetron sputtering in a mixture of Ar and $N_2$ gases (partial pressures of 79% and 21%, respectively). The thickness of the film was controlled by the sputtering time and verified using X-ray reflectometry. Here NbN was used because NbN SNSPD exhibited excellent performance at 2.1 K which could be realized using a commercial Gifford–McMahon cryocooler. The NbN film was then patterned into a meandered nanowire structure by electron beam lithography using a positive-tone polymethyl methacrylate electron-beam resist and was reactively etched in $CF_4$ plasma. Then, a 50-Ω-matched coplanar waveguide was formed using ultraviolet lithography and reactive ion etching.

Figure 2 shows the measurement setup of the SNSPD system. The optical module of this setup included a single-mode pulsed laser and two MMF attenuators connected with MMFs, which were used to generate 532 nm photons. The electronics of the SNSPD were composed of a bias circuit and a readout circuit, which were connected to the detector through a bias-tee and a coaxial cable with an impedance of 50 μm. The bias circuit included an isolated voltage source and a series resistor of 20 k, which formed a quasi-constant-current bias to the SNSPD via the DC arm of the bias-tee. The readout circuit included a low-noise wideband amplifier and a counter/SPC 150/oscilloscope connected to the SNSPD through the RF arm of the bias-tee. All electronics, except the SNSPD device, worked at room temperature. The SNSPDs were packaged into a copper block, such that a lensed multi-mode fiber could be directly aligned with the sensitive area of the SNSPD from the front side. The focus of the fiber was located on the center of the meandered nanowire, ensuring a maximal optical coupling from the fiber to the device. The package was mounted to the cold head of a two-stage Gifford–McMahon cryocooler with a working temperature of 2.100 ± 0.005 K. To measure the system detection efficiency (SDE), the single-mode fiber (SMF) laser module was directly connected to the MMF aligned to the SNSPD. Thus, only a small fraction of the MM fiber core was illuminated. No specific actions were taken to populate a large number of modes of the MMF, but this should not affect the measured DE since the diameter of the SNSPD was large enough to ensure a coupling of nearly unity, as described in the last paragraph, and our device showed a saturated DE behavior. We adopted an optical power meter to measure the power from the laser source and to calibrate two optical attenuators. Then the laser source and the optical attenuators were connected in series to function as a faint photon source with photon flux of $10^6$ photons/sec (~ -94.28 dBm), which was determined by the power of the laser source and the attenuation ratios of the attenuators. The SDE of the system was defined as (OPR − DCR)/PR, where OPR was the output pulse rate of the SNSPD, as measured using a photon counter; DCR was the dark count rate when the laser was blocked; and PR was the total photon rate input to the system. At each bias current, an automated shutter in a variable attenuator blocked the laser light, and dark counts were collected for 10 s using a commercial counter. The light was then unblocked and output photon counts were collected for another 10 s. Errors due to the calibration of the laser power were less than 3.5%, given by the power meter, and the laser power fluctuation was under 1.2%.

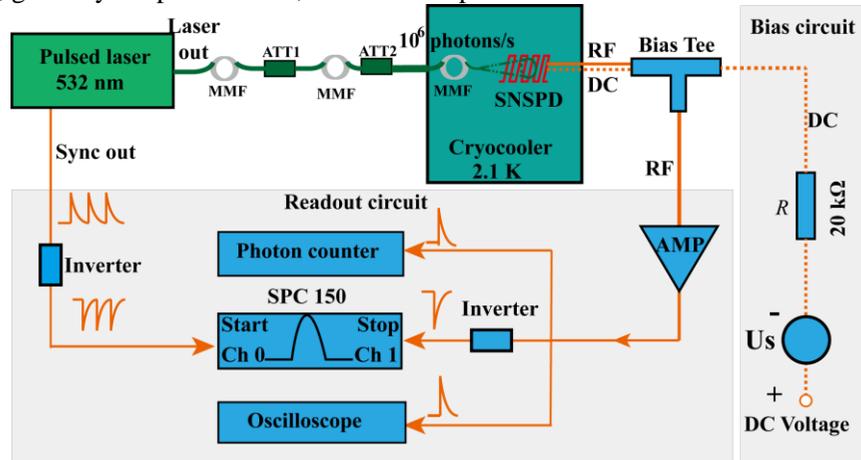

Fig. 2 System schematics for characterizing SNSPD. MMF: multimode fiber; ATT: variable attenuator; AMP: amplifier. The black dashed lines, solid lines with arrows, and red bold lines represent the DC path, RF path, and light path, respectively. The parallel solid and dashed lines represent the path for both DC and RF signals.

Figure 3(a) presents the I-V curves of our SNSPD indicating a switching current ($I_{sw}$) of 13.2 μA. The SDE and DCR relations of the bias current are shown in Fig. 3(b). A saturated DE behavior was registered at the bias current down to $0.7I_{sw}$ (9.0 μA) and the DE was approximately 75% at a DCR of <0.1 Hz. Since the MMF cannot maintain the polarization of the photons and the nanowire has a low polarization absorption ratio of 1.02 at the target wavelength of 532 nm according to our EM simulation, the system was almost insensitive to the polarization. Indeed, we measured the system DE polarization sensitivity by using polarization controller (Thorlabs: PLC-900) to tune the polarization of the photons, which gave a polarization sensitivity ratio of around 2.0%. The result is consistent with the simulation result. Note that the DCR was measured with MMF coupling to SNSPD and the connector at the room temperature was shielded. The extremely low DCR for MMF coupled SNSPDs was mainly due to the wilder nanowire, which was insensitive to infrared photons caused by room temperature blackbody radiation. Moreover, in comparison to previously reported MMF coupled SNSPDs for visible wavelengths [15], the background DCR was significantly decreased by two orders of magnitude. This is because the designed total refection band (460-580 nm) here was much smaller than that of the previous design (400-1,000 nm), which may have been beneficial to reduce the DCR caused by background blackbody radiation. The high DE SNSPD with a low DCR might benefit the observation of more distant and smaller targets in space.

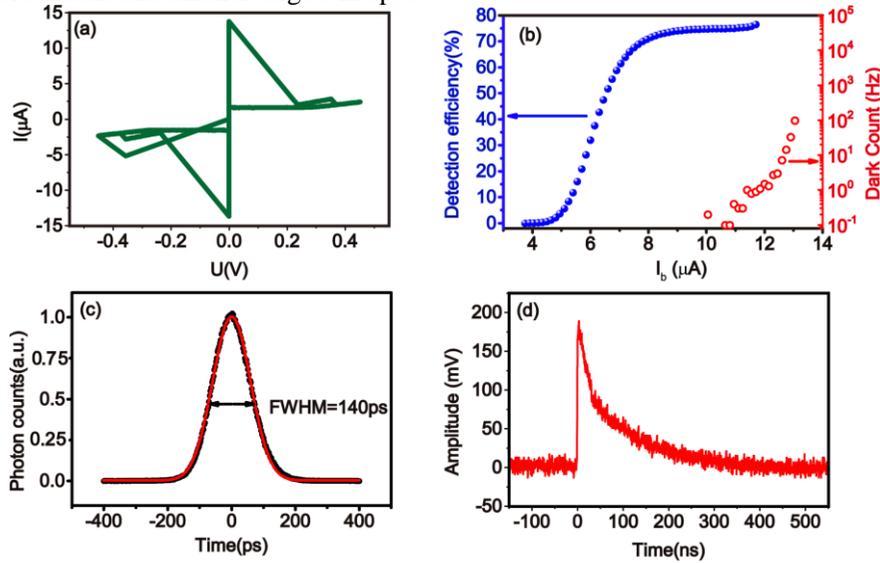

Fig. 3 (a) I-V curves of our SNSPD. (b)DE (left) and DCR (right) versus bias current. The DE is approximately 75% with DCR of <1Hz. (c) Histograms of the time-correlated photon counts measured at a wavelength of 1,550 nm. The red lines were the fitted curves using Gaussian distribution. (d) Oscilloscope persistence map of the response at a bias current of 10.0 μA.

The timing jitter of the SNSPD system was measured using the time-correlated single-photon counting (TCSPC) method[6]. To measure the jitter, one need an advanced femtosecond pulsed laser source with high timing resolution. Since we only have femtosecond pulsed laser (Calmar: FPL-01CFF, Pulse width: 0.5 ps，Jitter: <70 fs) at 1550 nm in lab, we measured the jitter of SNSPD at the wavelength of 1550 nm based on the assumption that the jitter is unchanged with the wavelength. Though the SDE at 1550 nm was around 1%, we can still acquire enough photon counts and obtain the jitter using TCSPC module (Becker & Hickl: SPC-150). At a bias current of 10.0 μA (DCR ~ 0.2 Hz), the timing jitter defined by the full-width at half-maximum (FWHM) value of the histogram was 140 ps. The corresponding standard deviation was about 59 ps which indicated a potential ranging precision of 8.9 mm. The jitter value was higher than the jitter of the SNSPDs with sensitive areas of 35 μm and 50 μm [15, 16] owing to the relatively smaller switching current and the larger kinetic inductance caused by the large sensitive area. Finally, the measured oscilloscope persistence trace of photon response pulse was presented in Fig. 4(d), which indicated a full recovery time of approximately 300 ns which is sufficient for the SLR application.

## 4. Satellite laser ranging

Using the designed SNSPD system, we carried out SLR experiments at the Sheshan Observatory (31°05′48″N121°11′24″E), of the Shanghai Astronomical Observatory, Chinese Academy of Sciences. Figure 4 shows the structure of the SLR system, including the satellite-prediction system, control system, pulse laser, laser-beam transmitting system, telescope mount tracking system, high-precision timing system, and return detection and receiving system. A bandpass filter (Center wavelength: 532 nm, Bandwidth: 2 nm) was used to filter out the background light before the returned photons were coupled into MMF. The apertures of the receiving and transmitting telescopes were 60 cm and 21 cm, respectively. The mount was of Alt-Azimuth type, and was directly driven by motors. The pointing accuracy of the telescope after star calibration is about 5". The Nd: YAG laser was a key instruments for the SLR with 1 mJ per pulse energy, 30 ps pulse width, 0.6 mrad divergence, 1 kHz repetition rate and 1 W mean power at a wavelength of 532nm.

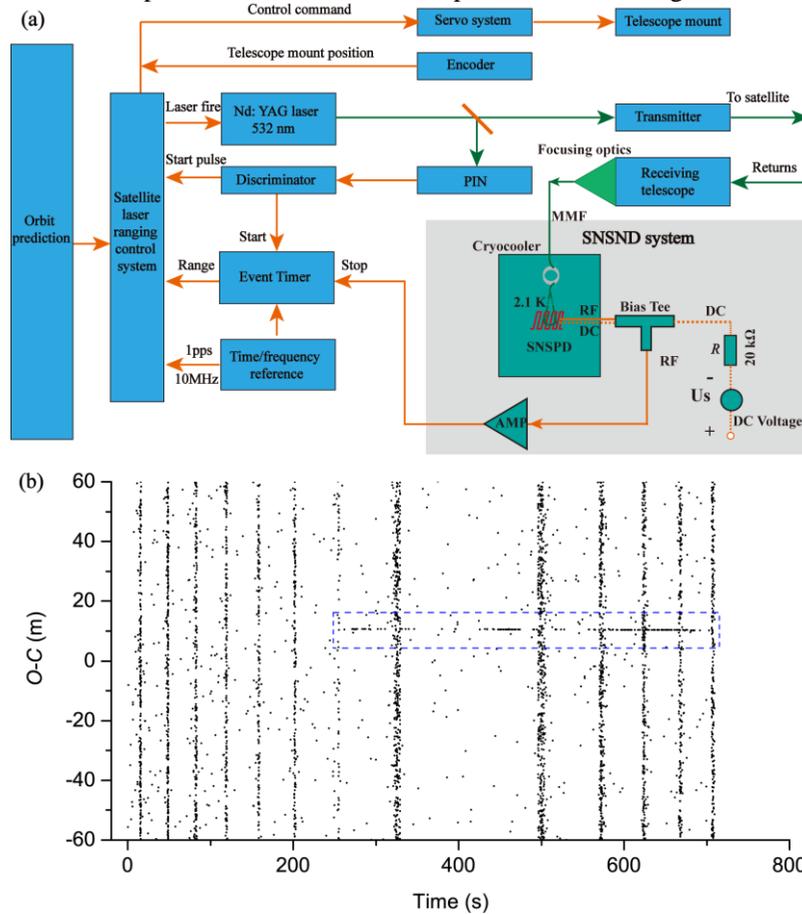

Fig. 4 (a) Schemetics of the SLR system using the deisgned SNSPD at the Sheshan Observatory (31°05′48″N121°11′24″E), Chinese Academy of Sciences. The system included a satellite-prediction system, control system, pulse laser, laser-beam transmitting system, telescope mount tracking system, high-precision timing system, receiving system and SNSPD system. (b) The ranging result. The horizontal axis is the time elapsed from the beginning of the tracking and the vertical axis is the range residual *O-C* between the observed and predicted values. The marked points in the dashed rectangular are returned photons' trace of the satellite LARES.

As a demonstration for the SNSPD system used in the SLR application, we chose LARES as the target satellite. LARES is a scientific satellite that was inserted in an orbit with 1,450 km of perigee, an inclination of 69.5° and reduced eccentricity. The distance from the satellite to the observation station was about 3,000 km. The ranging result is shown in Fig. 4(b), where the horizontal axis is the elapsed time from the beginning of the tracking. The vertical axis is the range residual *O-C* between the observed and predicted values. The points in the dashed rectangle are the returned photons' trace of LARES. The points on the vertical line are the laser returns due to atmospheric back-scattering and the other points are noise mainly due to the background environment. We extracted the ranging precision of about 8.0 mm by polynomial fitting the returned points, which is consistent with the expected 8.9 mm resolution estimated from the timing jitter. The deviation was caused by the fitting error due to the low detected photon counts. The precision was comparable with the same ranging system incorporated with APD($\varPhi$= 200 μm; DE >20 %, DCR= 8kHz @10Hz working frequency; Timing jitter = 94 ps.) which showed a ranging precision of about 10.0 mm. To further enhance the accuracy of the ranging, the timing jitter of the detector should be further improved. We also noticed that the detected returned photon count (137 in 420 s) was much less than the expectation, and corresponded to pulse detection efficiency (PDE) of 0.032%. The unexpectedly low count rate was mainly due to the coupling loss from the telescope to the MMF. Considering atmospheric seeing conditions, the static maximum coupling efficiency from our stationary telescope system to the MMF was roughly 30%. However, in practical ranging application, the receiving telescope and the MMF were kept moving in order to track the satellite. The system vibration and the asynchronous movement between the telescope and the MMF resulted in a much larger coupling loss even more than tens of dBm, which resulted in the low PDE value of 0.032%. It took the observation time of 420 seconds to obtain the ranging precision of about 8.0 mm. In comparison, the free-space coupled SPAD (Single Photon Avalanche Diode) was directly coupled to the receiving telescope and moved synchronously during the ranging experiment. A typical SPAD with the PDE of about 20% can be obtained and the observation time of a few seconds is enough to achieve a desired ranging precision based on the accumulated valid data. Thus, improvement of coupling from telescope to MMF in the near future may effectively improve the SLR performance using SNSPD.

## 5. Discussion

This study in this paper opens the road for SLR using SNSPDs. Especially for the future SLR at near-infrared wavelengths such as 1,064 nm or 1,550 nm, SNSPD will play an important role. Both the performance of the detector and the system can be further improved. It will be interesting to decrease the timing jitter to improve the ranging precision. One simple and effective approach is to increase the critical current by increasing the thickness of the nanowire appropriately. Furthermore, it is necessary to enhance the coupling from the telescope to the fiber-coupled SNSPD system, for example, by enlarging the active area of the SNSPD and applying the MMF fiber with a larger core diameter and a numerical aperture for better coupling of incident photons. Tapered optical fibers may be an alternative solution for efficient free space coupling of light to the nanowire region. Besides, the focusing optics used in the system need to be optimized carefully.

Based on the results above, we will study laser ranging at 1,064 nm in the near future. We expect that ranging systems incorporated with high performance SNSPDs can be applied to both SLR and space debris detection.

## 6. Conclusions

We designed, fabricated, and characterized NbN SNSPDs at 532 nm wavelength on a PC substrate with a sensitive area diameter of 42 μm. The PC structure effectively acted as a cavity to enhance the absorption of incident photons. The MMF coupled SNSPDs exhibited the maximum system DE of up 75% at DCR of <0.1 Hz as well as a timing jitter of 140 ps. SLR using the SNSPD exhibited a ranging precision of 8.0 mm for the satellite LARES (~3,000 km away from the ranging station), opening up new and interesting application of the SNSPD.

## Acknowledgments

This work was funded by the National Natural Science Foundation of China (Grant Nos. 61401441, and 61401443), Strategic Priority Research Program (B) of the Chinese Academy of Sciences (XDB04010200&XDB04020100).